\documentclass[preprint,prd]{revtex4} 
\font\titlefont=cmbx10 scaled \magstep3 

\begin{document}
\input{epsf} 

\begin{center}

{\titlefont Quantum Stress Tensor Fluctuations and their Physical Effects}


\vskip .7in
{L. H. Ford}
\vskip .1in
{Institute of Cosmology,
Department of Physics and Astronomy\\ 
         Tufts University, Medford, MA 02155\\
email: ford@cosmos.phy.tufts.edu}
\vskip .3in

{Chun-Hsien Wu}
\vskip .1in
{Institute of Physics,
Academia Sinica \\
Nankang, Taipei 11529 Taiwan\\
email: chunwu@phys.sinica.edu.tw}
\end{center}
\vskip .3in
\centerline{\bf Abstract}
\vskip .1in

\baselineskip=16pt
We summarize several aspects of recent work on quantum stress tensor 
fluctuations and their role in driving fluctuations of the
gravitational field. The role of correlations and anticorrelations is 
emphasized. We begin with a review of the properties of the stress
tensor correlation function. We next consider some illuminating examples of
non-gravitational effects of stress tensors fluctuations, specifically
fluctuations of the Casimir force and radiation pressure fluctuations.
We next discuss passive fluctuations of spacetime geometry and some
of their operational signatures. These include luminosity
fluctuations, line broadening, and angular blurring of a source
viewed through a fluctuating gravitational field. Finally, we discuss 
the possible role of quantum stress tensor fluctuations in the early 
universe, especially in inflation. The fluctuations of the expansion of
a congruence of comoving geodesics grows during the inflationary era,
due to non-cancellation of anticorrelations that would have occurred in
flat spacetime. This results in subsequent non-Gaussian density
perturbations and allows one to infer an upper bound on the duration
of inflation. This bound is consistent with adequate inflation to
solve the horizon and flatness problems.
\newpage


\baselineskip=14pt

\section{Introduction}

As is well-known, the classical stress tensor, $T_{\mu\nu}$, is both
the source of the gravitational field in general relativity theory,
and the quantity which describes stresses on material objects. In
quantum field theory, the stress tensor becomes an operator whose
expectation value is formally infinite, and needs to
be renormalized. In Minkowski spacetime, this is usually accomplished
by simply subtracting the vacuum expectation value, and replacing
the stress tensor operator by its normal-ordered version:
\begin{equation}
:T_{\mu\nu}: = T_{\mu\nu} - \langle T_{\mu\nu} \rangle_0 \,,
\end{equation}
where $\langle \; \rangle_0$ denotes an expectation value in the 
Minkowski vacuum state.
This amounts to defining the zero of energy density to be at the
vacuum level. This allows for states with local negative energy
densities, although the total energy must be non-negative, and the
regions of negative energy density are severely constrained by quantum
inequalities~\cite{Roman04}. In curved spacetime, the renormalization of 
$\langle T_{\mu\nu} \rangle$ is more complicated and involves renormalization
of the cosmological constant, Newton's constant, and the coefficients
of counterterms quadratic in the curvature. 

Our concern will not be with issues of renormalization of 
$\langle T_{\mu\nu} \rangle$ or with quantum violation of classical
energy conditions, but rather with fluctuations of the stress tensor
operator about its mean value. That there must be such fluctuations
in all realizable quantum states follows from the fact that these
states are never eigenstates of the stress tensor operator.

The gravitational effects of  stress tensor fluctuations have been
discussed by several authors in recent 
years~\cite{F82,KF93,HS98,H99,PH01,HV03,CH95,CCV,RV,Stochastic,Wu06}.
In this paper, we will discuss a selection of topics relating to
the basic character of stress tensor fluctuations, their role in
creating fluctuating forces on material bodies, and especially
their role in gravitational physics.

\section{The Stress Tensor Correlation Function}

The basic object which we will need to study in order to understand 
stress tensor fluctuations, will be the correlation function,
\begin{equation}
C_{\mu\nu\alpha\beta}(x,x') = 
\langle  T_{\mu\nu}(x)\, T_{\alpha\beta}(x') \rangle
- \langle  T_{\mu\nu}(x) \rangle \langle T_{\alpha\beta}(x') \rangle
\,.  \label{eq:corr_fnt}
\end{equation}
This object is independent of the choice of renormalization of 
$\langle T_{\mu\nu} \rangle$, as $C_{\mu\nu\alpha\beta}(x,x')$ is
unchanged if we shift the stress tensor by a c-number. It is, however,
singular in the coincidence limit $x' \rightarrow x$, even if 
$\langle T_{\mu\nu} \rangle$ is finite. It is often useful to
decompose the correlation function into three parts with differing
singularities. Here we consider only the case of Minkowski spacetime, 
but an analogous decomposition may be defined in curved spacetimes. 

The stress tensor for a free quantum field is a sum of terms, each
quadratic in field operators or derivatives of field
operators. Consider the stress tensor for a bosonic field, which is a
sum of terms of the form $T(x) = :\phi_1(x) \phi_2(x):$ Now consider
products of operators of the form of $T$. It may be shown using Wick's 
theorem that
\begin{equation}
T(x)\, T(x') = S_0 +S_1 + S_2 \,,
\end{equation}
where
\begin{equation}
S_0 = 
\langle \phi_1(x) \phi_1(x') \rangle_0 \langle \phi_2(x) \phi_2(x') \rangle_0 
+ \langle \phi_1(x) \phi_2(x') \rangle_0 \langle \phi_2(x) \phi_1(x') \rangle_0
\,,
\end{equation}

\begin{eqnarray}
S_1 = :\phi_1(x) \phi_1(x'):  \langle \phi_2(x) \phi_2(x') \rangle_0 +
:\phi_1(x) \phi_2(x'):  \langle \phi_2(x) \phi_1(x') \rangle_0 + \nonumber \\
:\phi_2(x) \phi_1(x'):  \langle \phi_1(x) \phi_2(x') \rangle_0 +
:\phi_1(x) \phi_2(x'):  \langle \phi_2(x) \phi_1(x') \rangle_0 \,, 
\label{eq:S1}
\end{eqnarray}
and
\begin{equation}
S_2 = :\phi_1(x) \phi_2(x) \phi_1(x') \phi_2(x'): \,.
\end{equation}
Thus the operator product $T(x)\, T(x')$ consists of a purely vacuum part
$S_0$, a fully normal-ordered part $S_2$, and a part $S_1$ which is a
cross term between the vacuum and normal-ordered parts. 

The same decomposition holds for the correlation function, which can
be written as a sum of normal-ordered, cross and vacuum terms:
\begin{equation}
C^{\mu\nu\alpha\beta}(x,x') = C^{\mu\nu\alpha\beta}_{NO}(x,x') +
C^{\mu\nu\alpha\beta}_{cross}(x,x') +
C^{\mu\nu\alpha\beta}_{vac}(x,x')\,.
\end{equation}
Here $C^{\mu\nu\alpha\beta}_{NO}(x,x')$ is state-dependent and finite
as $x' \rightarrow x$, the cross term is also state-dependent and
singular,
\begin{equation}
C^{\mu\nu\alpha\beta}_{cross}(x,x') \sim \frac{1}{(x-x')^4}\,,
\end{equation}
and the vacuum term is state-independent and singular
\begin{equation}
C^{\mu\nu\alpha\beta}_{vac}(x,x') \sim \frac{1}{(x-x')^8}\,.
\end{equation}

The singularities in the correlation function should not be a cause
for concern, as physical observables are integrals of the correlation
function, and can be defined by an integration by parts procedure. In
this sense, $C^{\mu\nu\alpha\beta}(x,x')$ is well-defined as a 
distribution. To illustrate the basic idea, consider the integral
\begin{equation}
\int_a^b \frac{f(x)}{(x-c)^n} \, dx \,,
\end{equation}
where $a < c < b$. We may use the identity
\begin{equation}
 \frac{1}{(x-c)^n} = 
(-1)^{n-1} (n-1)!\, \frac{d^{n}}{dx^{n}}\, \ln(x-c)  \, ,
\end{equation}
to first rewrite the integrand, and then integrate by parts to obtain
\begin{equation}
\int_a^b \frac{f(x)}{(x-c)^n} \, dx = -(n-1)! \int_a^b f^{(n)}(x)\,
\ln(x-c)\, dx \; +{\rm surface \, terms}  \,.
\end{equation}
The last integral contains only an integrable singularity, and
the surface terms are evaluated away from the singularity at $x=c$.
Thus if the function $f$ and its first $n$ derivatives are finite, then
the integral is well defined. An alternative approach is to use
dimensional regularization, in which case the integrals of
 $C^{\mu\nu\alpha\beta}(x,x')$ are finite in the limit of four
spacetime dimensions~\cite{FW04}.

\section{Fluctuations of Forces on Material Bodies}

Before turning to the main topic of this paper, it is informative to
give a brief summary of a closely related subject. Just as the
classical stress tensor may be used, for example, to compute electromagnetic
forces on dielectric bodies, the quantum stress tensor describes the
quantum fluctuations in these forces. An example is the Casimir force,
the mean value of which is given by a suitably defined expectation
value of $\langle T_{\mu\nu} \rangle$. This force is expected to
undergo fluctuations around this mean
value~\cite{Barton,Eberlein,JR,WKF02}. However these fluctuations 
are too small to be readily  observable. For example, in
Ref.~\cite{WKF02}, the fluctuations of the Casimir-Polder force on
a polarizable atom near a reflecting plate was calculated. It was
found that the atom undergoes Brownian motion in the sense that
its mean squared velocity shifts due to the presence of the plate.
The transverse component shifts by
 \begin{equation}
   \langle\Delta v_{x}^{2}\rangle
   =\frac{47}{768}\frac{\hbar^2 \,\alpha^{2}}{\pi^{4}m^{2}z^{8}}
    \label{eq:final_x} 
\end{equation}
and the longitudinal component by
\begin{equation}
   \langle\Delta v_{z}^{2}\rangle
   =-\frac{3787}{3840}\frac{\hbar^2 \,\alpha^{2}}{\pi^{4}m^{2}z^{8}}
    \label{eq:final_z} \, .
\end{equation}
Here $z$ is the distance to the plate, $m$ is the mass of the atom,
and $\alpha$ is its static polarizability in Lorentz-Heaviside units.
The negative sign in the longitudinal component seems to imply a 
reduction in the velocity dispersion of the wavepacket of a quantum
particle localized near the plate. Equations~(\ref{eq:final_x}) and
(\ref{eq:final_z}) represent a sum of a fully normal ordered (with
respect to the Minkowski vacuum) contribution, and a cross term.
However, in both cases, the dominant contribution is that of the
cross term. The effective temperature associated with transverse
component is
\begin{equation}
T_{eff} \approx 10^{-1} K \left(\frac{m_H}{m}\right)
\left(\frac{10^{-8} {\rm cm} }{z}\right)^8 
\left(\frac{\alpha}{\alpha_H}\right)^2\,,   
\end{equation}
where $m_H$ and $\alpha_H$ are the mass and static polarizability of atomic
hydrogen, respectively. Although the effect is small, it might be
observable if sufficiently small values of $z$ could be attained.

A second example of force fluctuation is the quantum fluctuation of
radiation pressure. This is expected to be a significant source of
noise in future generations of laser interferometer detectors of gravity
waves. For the case of light in a coherent state, this effect was first
analyzed by Caves~\cite{Caves1,Caves2}, using an approach based upon 
fluctuation in photon
numbers. It was studied by the present authors~\cite{WF01} using the quantum
stress tensor, where it was shown that the radiation pressure
fluctuations arise entirely from the cross term in the correlation function.
This follows from the fact that for a single mode coherent state 
$|z \rangle$,
\begin{equation}
\label{eq:zz}
\langle z|:T_{\mu \nu }T_{\rho \sigma }:|z\rangle =\langle z|:T_{\mu 
\nu }:|z\rangle \langle z|:T_{\rho \sigma }:|z\rangle \, ,
\end{equation}
and hence the fully normal ordered term vanishes, and the vacuum term
is independent of the state of the radiation field. If a free mirror of
mass $m$ is subjected to a laser beam moving in the $x$-direction for
a time $\tau$, then the variance in the mirror's velocity is
\begin{equation}
\label{eq:v2}
\langle \triangle v^{2}\rangle =\frac{1}{m^{2}}\int _{0}^{\tau 
}dt\int _{0}^{\tau }dt'\int _{A}da\int _{A}da'\langle 
T_{xx}(x)T_{xx}(x')\rangle _{cross}\, .
\end{equation}
Here $\int _{A}da$ denotes an integral over the area of the mirror.
If the laser beam has linear polarization in the $y$-direction, then
it may be shown that
\begin{equation}
\label{eq:Tcross2}
\langle T_{xx}(x)T_{xx}(x')\rangle _{cross}=\langle 
:B_{z}(x)B_{z}(x'):\rangle \langle B_{z}(x)B_{z}(x')\rangle _{0}\, ,
\end{equation}
where $\langle :B_{z}(x)B_{z}(x'):\rangle$ is a finite state-dependent
factor, and $\langle B_{z}(x)B_{z}(x')\rangle _{0}$ is the vacuum
magnetic field two-point function in the presence of the
mirror. Although the latter function is singular as $x' \rightarrow
x$, the integral in Eq.~(\ref{eq:v2}) is finite, and may be evaluated
to find
\begin{equation}
\label{eq:dv2_{s}t}
\langle \triangle v^{2}\rangle =4\, \frac{A\omega \rho }{m^{2}}\, 
\tau \, ,
\end{equation}
where $A$ is the mirror's area, $\omega$ is the angular frequency of
the laser beam, and $\rho$ is the energy density in the beam. This
agrees with Caves'~\cite{Caves1,Caves2} result using photon number
fluctuations.

\begin{figure}
\begin{center}
\leavevmode\epsfysize=8cm\epsffile{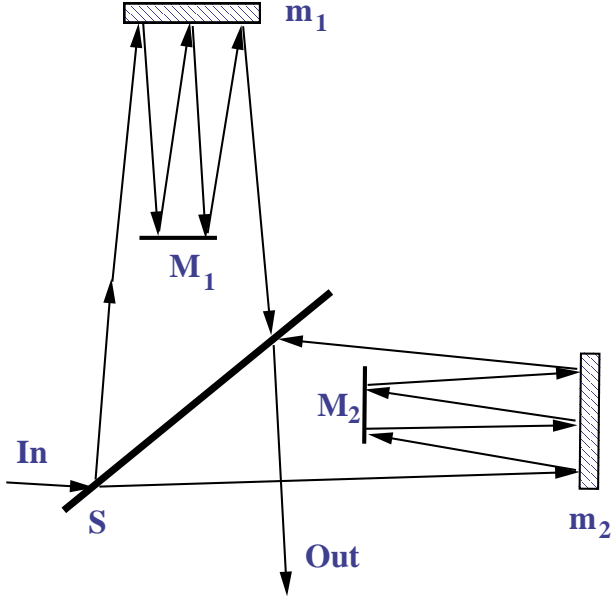} 
\end{center} 
  \caption{A Michelson interferometer with several bounces in each
    arm. The different illuminated spots on each mirror have
    correlated radiation pressure fluctuation.}
   \label{fig.interferometer}
\end{figure}

One remarkable aspect of the radiation pressure fluctuations is
that the fluctuations on different bounces in an interferometer
are correlated. A Michelson interferometer is illustrated in
Fig.~\ref{fig.interferometer}. The laser beam is split and
subsequently bounces $b$ times in each arm, illuminating $b$  different
spots on each mirror in the process. If the pressure fluctuations
at each spot were uncorrelated, the variance in a mirror's velocity
would be proportional to $b^2$, whereas in fact it is proportional to 
$b$. In a photon number approach these correlations come from the fact
that an fluctuation in photon number in a wavepacket is preserved as
the packet bounces in the interferometer. In the stress tensor
approach, the correlations of different spots in one arm are encoded in the
vacuum two-point function, $\langle B_{z}(x)B_{z}(x')\rangle _{0}$.
In the presence of mirrors, the usual lightcone singularity follows
the path of the beam in the interferometer.

\section{Anticorrelations and the Probability Distribution}

In the previous section, we saw an example in which stress tensor
fluctuations exhibit strong correlations. There are also examples 
of anticorrelations which we now discuss. First consider the case
of electric field fluctuations in the Minkowski vacuum. A charged
particle coupled to the fluctuating electric field should undergo
Brownian motion. However, if successive field fluctuations were
uncorrelated, then the particle's velocity should undergo a random
walk, and in the absence of damping we would have 
$\langle v^2 \rangle \propto t$. However, this would clearly violate
energy conservation, as there is no source of energy when the 
radiation field is in its vacuum state. The resolution of this
apparent paradox is that successive velocity fluctuations must be 
anticorrelated. The particle can acquire an energy $E$ from a
field fluctuation, but will lose it again on a timescale no longer than
$\hbar/E$. This effect can be analyzed clearly in the case where
one considers the effects of a reflecting plate, so that one is
calculating the change in $\langle v^2 \rangle$ due to the plate.
This was done in Ref.~\cite{YF04}. Consider a particle of mass
$m$ and charge $e$ which is at rest at $t=0$. At time $t$, the 
variance of the $i$-component of the velocity is 
\begin{equation}
\langle{\Delta v_i^2}\rangle={e^2\over m^2}\;\int_0^t\;\int_0^t\;
\langle{E}_i({\mathbf x},t_1)\;{E}_i({\mathbf x},t_2)\rangle_R\,dt_1
\, dt_2
 \; ,   \label{eq:lang2}
\end{equation}
where $\langle{E}_i({\mathbf x},t_1)\;{E}_i({\mathbf x},t_2)\rangle_R$
is the shift in the electric field correlation function due to the
plate. One finds that all components of $\langle{\Delta v_i^2}\rangle$
either vanish or approach a nonzero constant in the limit that
$t \rightarrow \infty$.  The result that 
$\langle{\Delta v_i^2}\rangle$ does not grow in time comes from the
fact that
\begin{equation}
\int_0^\infty\;
\langle{E}_i({\mathbf x},t_1)\;{E}_i({\mathbf x},t_2)\rangle_R\,dt_1
=0 \,,
\end{equation}
which is the mathematical expression of the anticorrelations.
Similar anticorrelations prevent the growth of the atom's velocity
dispersion in Eqs.~(\ref{eq:final_x}) and (\ref{eq:final_z}).

Analogous effects can be found in quantum stress tensor fluctuations
in the Minkowski vacuum state. Suppose that the energy density
along the worldline of an inertial observer is sampled using a
sampling function $ g(t,t_0)$, which is centered about $t=t_0$ and
has a characteristic width $a$. Defined the sampled energy density
by
\begin{equation} 
S(t_0)=\int_{-\infty}^{\infty} dt \, g(t,t_0) \, :T_{tt}(t):  \,, 
\label{eq:S} 
\end{equation}
Now perform two successive measurements, the first with the sampling
function $ g(t,0)$, and the second with the function $ g(t,t_0)$,
and then define the quantity
\begin{equation} 
\langle S(t_0) S(0) \rangle = \int_{-\infty}^{\infty} dt \,  
\int_{-\infty}^{\infty} \, dt' \, g(t,t_0) \, g(t',0)  
\, C(t,t') \, , 
\label{eq:SS}  
\end{equation}
where $C(t,t') = C_{tttt}(t,t')$ is the energy density correlation
function evaluated at coincident points in space, but distinct
times. The function $\langle S(t_0) S(0) \rangle$ describes the
correlation between the two successive measurements, the first
of the energy density in an interval around $t=0$, and the second  
in an interval around $t=t_0$. The explicit evaluation of the integrals
in Eq.~(\ref{eq:SS}) requires an integration by parts and is performed
in Ref.~\cite{FR05} for several sampling functions. The results
reveal that $\langle S(t_0) S(0) \rangle > 0$ for $t_0 \ll a$, but
then becomes negative as $t_0$ increases. It may undergo several
damped oscillations, but approaches zero for $t_0 \gg a$.
(See, for example Fig.~5 in Ref.~\cite{FR05}.) The
physical interpretation is this behavior is that nearly overlapping
measurements are positively correlated, but measurements taken
somewhat further apart in time become anticorrelated. There is an
equal amount of correlation and anticorrelation in the sense that
\begin{equation} 
\int_{0}^{\infty} dt_0 \,  \langle S(t_0) S(0) \rangle =
0\,. \label{eq:Sint}
\end{equation}
These results show that quantum fluctuation in the Minkowski vacuum
state exhibit strict anticorrelations as well as correlations. We will
see later that in curved spacetime it is possible to upset the
anticorrelations.

An issue related to the anticorrelations is that of the probability
distribution for stress tensor fluctuations. Let $\bar{T}$ be a
renormalized stress tensor component averaged over some spacetime
region. One would like to construct a probability function
$P(\bar{T})$ which gives the probability of finding a given value of
 $\bar{T}$ in a measurement. This is an unsolved
problem which is currently being investigated. A preliminary account
of this work was summarized in Ref.~\cite{F06}. The function
$P(\bar{T})$ is asymmetric, as typically odd moments such as
$\langle \bar{T}^3 \rangle$ are nonzero. This means that the
distribution cannot be Gaussian. Furthermore, it is expected that
$P(\bar{T})=0$ for $\bar{T} < T_{QI}$, where $T_{QI}$ is a negative
lower bound determined by quantum inequalities. However, there is no
upper bound. Because  $\langle \bar{T} \rangle=0$, the mean of the 
distribution is at zero, but the area to the left of  $\bar{T}=0$ 
exceeds that to the right. This means that if one makes a
measurement of the average energy density in a given region of
spacetime, the result will be negative more frequently than positive, 
but when positive the result typically has a greater magnitude.

\section{Passive Quantum Gravity}

In general, the spacetime geometry should be subjected both to
fluctuations in the intrinsic degrees of freedom of gravity, the active
fluctuations, as well as the effects of quantum stress tensor
fluctuations, the passive fluctuations. In this paper, we are
concerned only with the latter. Even without consideration of the
active fluctuations, we can have a restricted theory with only
the passive gravitational field fluctuations. Here we will outline
such a theory on a flat background, following Ref.~\cite{FW03}.
We consider only the contribution of the vacuum term in the
stress tensor correlation function. For the electromagnetic field, the
correlation function may be written as
\begin{eqnarray}
C_{(V)}^{\mu\nu\sigma\lambda}(x,x') &=&
4\, (\partial_{\mu }\partial_{\nu }D)\, 
(\partial_{\sigma }\partial _{\lambda }D)\,  \nonumber \\
&+& 2\, g_{\mu \nu }\, (\partial_{\sigma }\partial_{\alpha }D)\, 
(\partial_{\lambda}\partial^{\alpha }D)\,
+ 2\, g_{\sigma \lambda}\, (\partial_{\mu }\partial_{\alpha }D)\, 
(\partial_{\nu}\partial^{\alpha }D)\,                 \nonumber \\
&-& 2\, g_{\mu \sigma }\, (\partial_{\nu }\partial_{\alpha }D)\, 
(\partial_{\lambda}\partial^{\alpha }D)\,
- 2\, g_{\nu \sigma }\, (\partial_{\mu }\partial_{\alpha }D)\, 
(\partial_{\lambda}\partial^{\alpha }D)\,              \nonumber \\
&-& 2\, g_{\nu \lambda }\, (\partial_{\mu }\partial_{\alpha }D)\, 
(\partial_{\sigma }\partial^{\alpha }D)\,
- 2\, g_{\mu \lambda }\, (\partial_{\nu }\partial_{\alpha }D)\, 
(\partial_{\sigma }\partial^{\alpha }D)\,            \nonumber \\
&+&(g_{\mu \sigma }g_{\nu \lambda }+g_{\nu \sigma }g_{\mu \lambda }-
 g_{\mu \nu }g_{\sigma \lambda })\,(\partial_{\rho }\partial_{\alpha }D)\, 
(\partial^{\rho}\partial^{\alpha }D) \,, \label{eq:CV}
\end{eqnarray}
where
\begin{equation}
D = D(x-x') = \frac{1}{4 \pi^2 (x-x')^2} \label{eq:scalar_2pt}
\end{equation}
is the Hadamard (symmetric two-point) function for the massless scalar field.
A similar result for the case of the scalar field has been given by
Martin and Verdaguer~\cite{MV99}.  Let $h_{\mu \nu }$
be a metric perturbation due to the stress tensor $T_{\mu \nu }$. In a
gauge in which $\partial_{\nu }{h}^{\mu \nu }= h^\mu_\mu =0$, the
linearized Einstein equation becomes
\begin{equation}
\partial_\alpha \partial^\alpha {h}_{\mu \nu }=-16\pi T_{\mu \nu }\,
\end{equation}
in units in which $G =1$, where $G$ is Newton's constant.
One may solve this equation as an integral involving the retarded
Green's function, and use the result to construct the metric
correlation function $\langle h^{\mu \nu }(x)h^{\rho \sigma
}(x')\rangle$. Remarkably, this correlation function can be expressed
as a local function,
\begin{eqnarray}
\langle h_{\mu \nu }(x)h_{\sigma \lambda}(x')\rangle  & = & 
-\frac{1}{60\, \pi^2}\Bigl[
 4\, \partial_{\mu }\partial_{\nu }\partial_{\sigma }\partial_{\lambda }\, S
+ 2 \,(g_{\mu \nu }\, \partial_{\sigma }\partial_{\lambda }
   + g_{\sigma \lambda}\, \partial_{\mu }\partial_{\nu} )\, 
\partial_\alpha \partial^\alpha S \nonumber \\
&-& 3\, (g_{\mu \sigma }\, \partial_{\nu }\partial_{\lambda} +
     g_{\mu \lambda }\, \partial_{\nu }\partial_{\sigma} +
     g_{\nu \sigma }\, \partial_{\mu }\partial_{\lambda} +
     g_{\nu \lambda }\, \partial_{\mu }\partial_{\sigma})\, 
\partial_\alpha \partial^\alpha S \nonumber \\
&+& 3\, (g_{\mu \sigma }g_{\nu \lambda }+
                              g_{\nu \sigma }g_{\mu \lambda })\,
(\partial_\alpha \partial^\alpha)^2 S
- 2\, g_{\mu \nu }g_{\sigma \lambda }\,
(\partial_\alpha \partial^\alpha)^2 S \Bigr]\,,    \label{eq:h-h-2p} 
\end{eqnarray}
where
\begin{equation}
S = \ln^{2}[\mu^{2}(x-x')^{2}]\, , 
\end{equation}
and $\mu$ is an arbitrary constant. The significance of
Eq.~(\ref{eq:h-h-2p}) is that expresses the metric correlation function
as a sum of total derivatives. This allows us to readily evaluate
observable quantities associated with metric fluctuations using
integration by parts, as these quantities are expressible as integrals
involving the correlation function.

\section{Some Physical Effects of Passive Geometry Fluctuations}

In this section, we will summarize three physical phenomena which
in principle could be caused by quantum geometry fluctuations near
a flat background.

\subsection{Luminosity Fluctuations} 
The image of a distant source viewed through a fluctuating medium
will undergo variations in apparent luminosity. This effect is the
cause of the familiar ``twinkling'' of stars viewed thought the earth's 
atmosphere. In principle, the same effect will arise due to geometry
fluctuations. A quantitative discussion of this effect was given in
Ref.~\cite{BF04a}. The basic idea is that luminosity fluctuations are
related to the expansion parameter, $\theta$, of a bundle of
geodesics.
If $k^\mu$ is the tangent vector to the geodesics, $\theta$ is defined
by $\theta = k^\mu_{;\mu}$, or equivalently as the logarithmic
derivative of the cross sectional area of the bundle,
\begin{equation}
{\theta} = {d \log ({A}/A_{0}) \over d\lambda}\, ,
\end{equation}
where $\lambda$ is the affine parameter for the geodesics. The
expansion satisfies the Raychaudhuri equation, 
\begin{equation}
\frac{d \theta}{d \lambda} = - R_{\mu\nu} k^\mu k^\nu - a\, \theta^2
-\sigma_{\mu\nu} \sigma^{\mu\nu} + \omega_{\mu\nu}  \omega^{\mu\nu} \,.
                                                 \label{eq:ray}
\end{equation}
Here $\sigma^{\mu\nu}$ is the shear and $\omega^{\mu\nu}$ is the vorticity
of the congruence. The constant $a=1/2$ for null geodesics, and $a=1/3$ for 
timelike geodesics. The use of the Raychaudhuri equation as a Langevin
equation for fluctuating spacetimes was proposed by Moffat~\cite{Moffat}.
We are interested in situations where shear, vorticity and $\theta^2$
may be neglected, in which case
\begin{equation}
\frac{d \theta}{d \lambda}  = - R_{\mu\nu} k^\mu k^\nu \, ,
\end{equation}
where $ R_{\mu\nu}$ is determined by the fluctuating stress tensor by
\begin{equation}
R_{\mu\nu} = 8 \pi \left(T_{\mu\nu} - 
     \frac{1}{2} g_{\mu\nu}\, T^\rho_\rho \right) \,.  \label{eq:Einstein}
\end{equation}
For photons on a nearly flat background, we may take the affine
parameter to be the coordinate time in our frame. The luminosity
fluctuations may be shown to be an integral of the expansion
correlation function
\begin{equation}
\left\langle\;\left(\Delta L \over L\right)^{2}\;\right\rangle =
  \int_{0}^{s}
  \int_{0}^{s}dt'\; dt'' \; 
\left[\langle \theta(t')\;\theta(t'') \rangle
- \langle \theta(t')\rangle\;\langle \theta(t'') \rangle \right] \, ,
\end{equation}
where $s$ is the flight distance.
However, in many cases~\cite{BF04a} this integral is approximately
\begin{equation}
\left\langle\;\left(\Delta L \over L\right)^{2}\;\right\rangle \propto
  \;
  s^2 \left\langle\; (\Delta \theta)^2\;\right\rangle.\label{eq:dellum}
\end{equation}    
The variance of $\theta$ may be computed as an integral of the Ricci
tensor correlation function 
\begin{equation}
\langle \theta^2 \rangle- \langle \theta \rangle^2 =
\langle (\Delta \theta)^2 \rangle=
\int_0^{\lambda_0} d\lambda \int_0^{\lambda_0} d\lambda' \,
 K_{\mu \nu \alpha \beta}(\lambda,\lambda')\, k^\mu (\lambda) k^\nu (\lambda)
\, k^\alpha(\lambda') k^\beta(\lambda') \,, \label{eq:var}
\end{equation}
where $ K_{\mu \nu \alpha \beta}$ is the Ricci tensor correlation
function, which is algebraically related to the stress tensor
correlation function, Eq.~(\ref{eq:corr_fnt}). 

The luminosity fluctuations caused by the stress tensor fluctuations
in a thermal bath of scalar particles is~\cite{BF04a}
\begin{equation}
\left({\Delta L \over L}\right)_{rms} = 0.02\, 
\left(\frac{s}{10^{28} {\rm cm}} \right)^\frac{3}{2}
\, \left(\frac{T}{10^6 K} \right)^\frac{7}{2}  =
10^{-3}\, \left(\frac{s}{10^{6} {\rm km}} \right)^\frac{3}{2}
\, \left(\frac{T}{1 {\rm GeV}} \right)^\frac{7}{2}\, . \label{eq:thermal}
\end{equation}
This result holds when the wavelength of the radiation from the source
is short compared to the typical wavelength of the particles in the
thermal bath. It shows that although the effects of spacetime geometry
fluctuations are small, there are conceivable circumstances in which
they are nonzero.

\subsection{Line Broadening and Angular Blurring}

In addition to luminosity fluctuations, there are other possible
effects of a fluctuating gravitational field upon light propagation.
These include the broadening of spectral lines and the angular blurring
of images. These two effect may both be expressed in terms of
integrals of the Riemann tensor correlation function~\cite{RF06}.
This geometric construction relies upon the fact that the change
in a vector, when parallel transported around a closed path in curved
spacetime, is the integral of the Riemann tensor over the enclosed
surface. Consider two observers who are initially at rest with respect
to one another and have 4-velocity $t^\mu$. Suppose that two
successive photons are sent from one observer to the other. If there
is classical gravitational field present, then there will be a
fractional frequency shift of $\xi = \Delta \omega/\omega$.  If
the gravitational field fluctuates, then this fractional shift has
a variance of
\begin{equation}
  \delta\xi^2 = \langle(\Delta\xi)^2\rangle -
  \langle\Delta\xi\rangle^2 =
  \int da \int da' \,
  C_{\alpha\beta\mu\nu\,\gamma\delta\rho\sigma}(x,x')
  t^{\alpha}k^{\beta}t^{\mu}k^{\nu}t^{\gamma}k^{\delta}t^{\rho}k^{\sigma}\,,
                    \label{eq:delta_xi}
\end{equation}
where $k^\mu$ is the photon wavevector and 
$C_{\alpha\beta\mu\nu\,\gamma\delta\rho\sigma}(x,x')$ is the 
Riemann tensor correlation function:
\begin{equation}\label{CorFunction}
 C_{\alpha\beta\mu\nu\,\gamma\delta\rho\sigma}(x,x') = \langle
 R_{\alpha\beta\mu\nu}(x)
 R_{\gamma\delta\rho\sigma}(x')\rangle -
 \langle R_{\alpha\beta\mu\nu}(x)\rangle
 \langle R_{\gamma\delta\rho\sigma}(x')\rangle\,.
\end{equation}
The fluctuation in the angle of the source in a direction specified by
a unit spacelike vector $s^\mu$ is given by a similar expression, 
\begin{equation}
 \delta\Theta^2 = \langle(\Delta\Theta)^2\rangle -
 \langle\Delta\Theta\rangle^2 = \int da\int da' \,
 C_{\alpha\beta\mu\nu \, \gamma\delta\rho\sigma}(x,x')
 s^{\alpha}k^{\beta}t^{\mu}k^{\nu}s^{\gamma}k^{\delta}t^{\rho}k^{\sigma}\,.
                                    \label{eq:delta_theta}
\end{equation}

There is a crucial difference between luminosity fluctuations on the one
hand and both line broadening and angular blurring on the other. The
expansion fluctuations, and hence the luminosity fluctuations, are
determined by the Ricci tensor correlation function, and hence arise
in leading order only for passive geometry fluctuations. The other two 
effects, being determined by the Riemann tensor fluctuations, can
occur for active fluctuations as well. However, in many cases of
passive fluctuations, all three effects are of the same order. Thus
for a thermal bath, Eq.~(\ref{eq:thermal}) also gives an estimate of
the magnitude of the line broadening and angular blurring effects.

\subsection{Black Hole Fluctuations}

Fluctuations in both the Hawking radiation and in black hole horizons
have been discussed by several authors. In particular, fluctuations
in the outgoing flux at infinity were calculated in Ref.~\cite{WF99}.
This fluctuation, in dimensionless terms, is of order unity. This can
be understood by recalling that a radiating black hole emits on
average one particle in a time of order $M$. Thus the variance in this
number is also of order unity. The calculation of the mass
fluctuations of evaporating black holes requires consideration of the
backreaction, as recently noted by Hu and Roura~\cite{HR07}. This
leads to a larger result than the estimate given in Ref.~\cite{WF99}
because a positive fluctuation in the outgoing radiation causes the
black hole's mass to decrease and hence for it to radiate more
rapidly.

Fluctuation of the horizon are more difficult to calculate, and are
a topic of ongoing research~\cite{HR07,FS97,BFP00}.

\section{Stress Tensor Fluctuations in Inflation}

In this section, we will summarize the results of Ref.~\cite{WNF07},
where the effect of electromagnetic stress tensor fluctuations in
inflation were studied. Take the spacetime to be a spatially flat
Robertson-Walker universe
\begin{equation}
ds^2 = -dt^2 +a^2(t) (dx^2+dy^2+dz^2) \,. \label{eq:RW}
\end{equation}
In a general spacetime, the conservation law for a perfect fluid can 
be written as~\cite{Hawking66,Olson}
\begin{equation}
\dot{\rho}+(\rho+p)\,\theta =0 \,,
  \label{eq:conservation} 
\end{equation}
where $\theta$ is the expansion of the congruence of worldlines for
the  observers who measure the energy density $\rho$ and pressure $p$.
In the case of the unperturbed Robertson-Walker universe,
\begin{equation}
\theta = \theta_{0}=3\frac{\dot{a}}{a}\,,
   \label{eq:theta0}
\end{equation}
leading to the usual form of the conservation law. Our interest is
in fluctuations of  $\theta$, which will in turn lead to fluctuations
in $\rho$.  

Consider a conformally invariant field, such as the electromagnetic
field, and assume that the shear and vorticity terms in the
Raychaudhuri equation can be neglected, but retain the $\theta^2$
term. It may be shown that the expansion correlation function can
be expressed as an integral of the stress tensor correlation function
as
\begin{equation}
\langle \theta(t_1)\, \theta(t_2) \rangle - 
\langle \theta(t_1)\rangle \langle  \theta(t_2) \rangle = (8\pi)^2\,
a^{-2}(t_1)\,a^{-2}(t_2)\, \int_{t_{0}}^{t_{1}}dt\, a^{2}(t)
\int_{t_{0}}^{t_{2}}dt'\, a^{2}(t')\, C_{tttt}\,.
   \label{eq:theta_corr}
\end{equation}
The energy density correlation function can be obtained by a conformal
transformation from flat spacetime:
\begin{equation}
C_{ttt't'}(x,x')= a^{-4}(t)\,a^{-4}(t')\,{\cal E}\,,
\end{equation}
where ${\cal E}$ is the flat space vacuum energy density correlation
function. It is convenient to convert from comoving time $t$ to
conformal time $\eta$, where $dt=a\, d\eta$. The $\theta$-correlation
function now becomes
\begin{equation}
\langle \theta(\eta_1)\, \theta(\eta_2) \rangle - 
\langle \theta(\eta_1)\rangle \langle  \theta(\eta_2) \rangle =
\frac{(8\,\pi)^2}{\ a^{2}(\eta_1)\,a^{2}(\eta_2)}\; 
\int_{\eta_{0}}^{\eta_{1}}
\frac{d\eta}{a(\eta)}
\int_{\eta_{0}}^{\eta_{2}}
\frac{d\eta'}{a(\eta')}\; {\cal E}(\Delta\eta,r) \,,
   \label{eq:theta_corr2}
\end{equation}
where $\Delta\eta = \eta -\eta'$ and $r = |\mathbf{x} - \mathbf{x'}|$ is the
coordinate space separation of the pair of points at which $\theta$ is
measured. Here we  assume that the $\theta$-fluctuations vanish
at  $\eta = \eta_0$. For the electromagnetic field
\begin{equation}
{\cal E}_{em}= \frac{(r^{2}+3\Delta\eta^{2})^{2}}
{4\pi^{4}(r^{2}-\Delta\eta^{2})^{6}}\,.
\label{eq:EM_corr}
\end{equation}

Consider inflation followed by a radiation-dominated universe. We
take the scale factor to be
\begin{equation}
a(\eta) =  \frac{1}{1 - H \eta}\,,  \qquad \eta_0 \leq \eta \leq 0 \,, 
\end{equation}
and 
\begin{equation}
a(\eta) = 1+ H\,\eta \,,  \qquad  \eta \geq 0  \,.
 \label{eq:a_rad1}
\end{equation}
Thus reheating occurs on the $\eta=0$ surface. 
If we assume that the perfect fluid has the
equation of state $p=w\rho$, then the conservation law yields the
density fluctuations in the post-inflationary era:
\begin{equation}
\left\langle\left(\frac{\delta\rho}{\rho}\right)^{2}\right\rangle=
(8\,\pi)^2\; (1+w)^{2}\int_0^{\eta_{s}}\frac{d\eta_1}{a(\eta_1)}
\int_0^{\eta_{s}} \frac{d\eta_2}{a(\eta_2)} \,
[\langle \theta(\eta_1)\, \theta(\eta_2) \rangle - 
\langle \theta(\eta_1)\rangle \langle  \theta(\eta_2) \rangle] \,.
\label{eq:del_rho}
\end{equation} 
Here the integrations begin at $\eta =0$ because the classical fluid
is assumed to be created then. The power spectrum of density
perturbations, $P_{k}$, is defined by 
\begin{equation}
\left\langle\left(\frac{\delta\rho}{\rho}\right)^{2}\right\rangle  =
\int d^{3}k\, {\rm e}^{i\,\mathbf{k}\cdot\Delta\mathbf{x}}P_{k}(\eta_{s})\,.
\label{eq:power}
\end{equation}

Remarkably, one finds that the effect of the quantum stress tensor
fluctuations upon the density perturbations grows as the duration of
inflation increases. For $H|\eta_0| \gg 1$, one finds that the
variance of the expansion at the end of inflation is given by
\begin{equation}
 \langle  (\Delta\theta)^2 \rangle =
  \langle \theta(0)\, \theta(0) \rangle - 
\langle \theta(0)\rangle \langle  \theta(0) \rangle
\approx\frac{8 H^{2}|\eta_{0}|^{2}}{5\pi^{2}r^{6}} \,.
\label{eq:F0_asy}
\end{equation}
This result reveals that the expansion fluctuations grow with
 the duration of
inflation. In the flat space limit, $H \rightarrow 0$, there would be
no growth. We can understand the lack of growth in flat spacetime as
being due to anticorrelations. The growth in deSitter spacetime means
that the cancellations that would have occurred in flat spacetime have
been upset. This is implemented by the factors of $1/a$ which appear
in the integrand of Eq.~(\ref{eq:theta_corr2}), and reflect the time
dependent geometry of an expanding universe. 
   
The power spectrum of density perturbations also grows as
$|\eta_0|^2$,
\begin{equation}
P_{k}(\eta_{s})\approx
\frac{32 |\eta_{0}|^{2}k^{3}}{15\,\pi}\, 
\ln^2[a(\eta_s)]\,(1+w)^{2}\,\ell_{p}^{4}\,,
\label{eq:power3}
\end{equation}
where $\ell_{p}$  is the Planck length. If we consider the effects of
the modes within a finite bandwidth, $\Delta k \approx k$, then we 
find an estimate of the associated density perturbations
\begin{equation}
\left(\frac{\delta\rho}{\rho}\right)_{rms} \approx 
\ell_{p}^{2}\, |\eta_{0}|\, k^{3}\,  
\ln[a(\eta_s)]   \,.
\label{eq:power_estimate}
\end{equation}
Here $1/a(\eta_s)$ is the redshift factor between reheating and the
last scattering surface, so that
\begin{equation}
a(\eta_s) \approx \frac{E_R}{ {\rm 1 eV}} \,,
\end{equation}
where $E_R$ is the reheating energy scale. Note that $k = a_{\rm now}\, k_P$. 
Let $k_P = 2 \pi/\lambda$ correspond to the typical intergalactic separation
today, $\lambda \approx  2{\rm Mpc}$, or $k_P \approx 10^{-24} {\rm
  cm^{-1}}$. On this scale, we must have
\begin{equation}
\left(\frac{\delta\rho}{\rho}\right)_{rms} < 10^{-4} \,,
\end{equation}
which leads to an upper bound on the duration of inflation of
\begin{equation}
H\, |\eta_{0}| < 10^{79}\, \left(\frac{10^{12} {\rm GeV}}{E_R}\right) \,.
   \label{eq:bound1}
\end{equation}
This bound is considerably greater than the amount of inflation needed
to solve the horizon and flatness problems, $H\, |\eta_{0}| >
10^{23}$. The density perturbations described by
Eq.~(\ref{eq:power_estimate}) can be understood as arising from a
kinematic effect. The $\theta$ fluctuations which accumulate during
inflation cause differential expansion rates at the reheating surface,
$\eta = 0$, which are transmitted through this surface to the radiation  
dominated universe. The density variations then arise from  differential
redshifting.

It is possible to obtain a stronger bound by examining the dynamics
of the inflaton field, $\varphi$. The expansion $\theta$ appears in the 
inflaton equation of motion, so that $\theta$ fluctuations create
$\varphi$ fluctuations. these are in addition to the intrinsic quantum
fluctuations of $\varphi$, which are usually considered. In
Ref.~\cite{WNF07}, it was shown that the the expansion induced density
perturbations are
\begin{equation}
\left(\frac{\delta\rho}{\rho}\right)_{rms}
\approx \sqrt{k^3\, P_{k}} \approx 
10^{-2} \frac{\ell_{p}^2\, H|\eta_{0}|\, k} {t_{R}} \,.
\label{eq:power_estimate2}
\end{equation}
With the same choice of scale as above, we find
\begin{equation}
H\, |\eta_{0}| <
10^{45}\, \left(\frac{10^{12} {\rm GeV}}{E_R}\right)^3  \,.
   \label{eq:bound2}
\end{equation}
 This is considerably more restrictive than  Eq.~(\ref{eq:bound1}), but is 
still compatible with adequate inflation to solve the horizon problem.

\section{Summary}
 
We have seen that quantum stress tensor fluctuations can have several
physical effects. Because the stress tensor describes the forces which
the electromagnetic field exerts on a material body, stress tensor
fluctuations result in effects such as Casimir force fluctuations and
radiation pressure fluctuations. These systems can be useful analog
models for gravitational field fluctuations and are potentially
amenable to study in the laboratory. 

A crucial feature of  stress tensor fluctuations is the presence of
subtle correlations and anticorrelations. In flat spacetime, the 
anticorrelations often lead to the exact cancellation of the effects
of the fluctuations, as in the case illustrated in
Eq.~(\ref{eq:Sint}). However, in curved spacetime these cancellations
can be altered, as in the case of deSitter spacetime where the
variance of the expansion parameter $\theta$ grows in time.

Quantum stress tensor fluctuations lead to passive fluctuations of the
gravitational field, which are in addition to the active fluctuations
coming from the quantization of the gravitational field itself. These 
fluctuations around a flat background are described by a metric
correlation function, which for the electromagnetic field in the
vacuum state is given by Eq.~(\ref{eq:h-h-2p}). In more general quantum
states for the matter field, one can have various physical phenomena
produce by gravitational field fluctuations, These include effects
on the propagation of signals through the fluctuating background,
such as luminosity fluctuations, line broadening, and angular blurring
of images. All of these effects can be described in terms of integrals
of the Riemann tensor correlation function, and for passive spacetime
fluctuations, in terms of the stress tensor correlation function.
Although the stress tensor correlation function is formally singular
in the limit of coincident points, it is a well defined distribution
in the sense that its integrals can be evaluated using integration
by parts. 

Passive geometry fluctuations can play a role in both black hole
evaporation and in the early universe. We have seen how the effects of
the fluctuations of the electromagnetic field can cause the expansion
of the comoving geodesics in an inflationary universe to undergo
growing fluctuations. These expansion fluctuations in turn lead
to density perturbations in the post-inflationary universe that are
potentially observable. These perturbations have a non-scale invariant
and non-Gaussian nature, and must hence constitute at most a small
fraction of the primordial density fluctuations. This allows one to
place constraints on the duration of inflationary expansion.

\begin{acknowledgments}
We are grateful to numerous colleagues for valuable discussions on
stress tensor fluctuations. Among them are Jake Borgman, Chris
Fewster, Bei-Lok Hu,
Chung-I Kuo, Kin-Wang Ng, Tom Roman, Albert Roura, Robert Thompson,
Enric Verdaguer, and Richard Woodard.  
  This work was supported in part by the National Science Foundation
under Grant No. PHY-0555754. 
\end{acknowledgments}


\begin{thebibliography}{}

\bibitem{Roman04} See, for example, the following article and
  references therein: T.A. Roman, in {\it Proceedings of the Tenth
  Marcel Grossmann Meeting}. M. Novello, S. Perez-Bergliaffia, and
  R. Ruffini, eds. (World Scientific, Singapore, 2005) Part C,  p1909;
  gr-qc/0409090.  


\bibitem{F82} L.H. Ford, Ann. Phys (NY) {\bf 144}, 238 (1982).

\bibitem{KF93} Chung-I Kuo and L.H. Ford, Phys. Rev. D {\bf 47}, 4510 (1993),
gr-qc/9304008.

\bibitem{HS98}  B.L. Hu and K. Shiokawa, Phys. Rev. D  {\bf 57}, 3474 (1998),
 gr-qc/9708023.

\bibitem{H99} B.L. Hu, Int .J. Theor. Phys. {\bf 38}, 2987 (1999),
gr-qc/9902064.

\bibitem{PH01} N.G. Phillips and B.L. Hu, Phys. Rev. D  {\bf 63}, 104001
(2001), gr-qc/0010019.

\bibitem{HV03} B.L. Hu and E. Verdaguer, Class. Quant. Grav. {\bf 20},
R1 (2003),  gr-qc/0211090.

\bibitem{CH95} E. Calzetta  and B.L. Hu,  Phys. Rev. D  {\bf 49}, 
6636 (1994), gr-qc/9312036;  
Phys. Rev. D {\bf 52}, 6770 (1995), gr-qc//9505046.

\bibitem{CCV}  E. Calzetta, A. Campos, and E. Verdaguer, Phys. Rev. D 
 {\bf 56},
2163 (1997),  gr-qc/9704010.

\bibitem{RV} R. Martin and E. Verdaguer, Phys. Lett. B {\bf 465}, 113 (1999);
 Phys. Rev. D  {\bf 61}, 124024 (2000),  gr-qc/0001098.

\bibitem{Stochastic} B.L. Hu and E. Verdaguer, Living Rev. Rel. {\bf 7},
3 (2004), gr-qc/0307032.

\bibitem{Wu06} C.-H. Wu, K.-W. Ng, W. Lee, D.-S. Lee, and Y.-Y. Charng,
JCAP {\bf 2}, 6 (2007), astro-ph/0604292.


\bibitem{FW04} L.H. Ford and R.P. Woodard, Class. Quant. Grav. {\bf 22}, 
1637 (2005), gr-qc/0411003.

\bibitem{Barton}G. Barton, J. Phys. A {\bf 24}, 991 (1991);
{\bf 24}, 5563 (1991).

\bibitem{Eberlein} C. Eberlein, J. Phys. {\bf A 25}, 3015 (1992);
{\bf A 25}, 3039 (1992).

\bibitem{JR} M.T. Jaekel and S. Reynaud, Quantum Opt. {\bf 4}, 39 
(1992);
J. Phys. I France {\bf 2}, 149 (1992); {\bf 3}, 1 (1993); {\bf 3}, 
339 (1993).

\bibitem{WKF02} C.-H. Wu,  Chung-I Kuo,  and L.H. Ford, Phys. Rev. A
 {\bf 65}, 062102 (2002); quant-ph/0112056. 

\bibitem{Caves1}C. M. Caves, Phy. Rev. Lett. \textbf{45}, 75 (1980).

\bibitem{Caves2}C. M. Caves, Phys. Rev. D \textbf{23}, 1693 (1981).

\bibitem{WF01} C.-H. Wu and L.H. Ford, Phys. Rev. D {\bf 64}, 045010 (2001). 


\bibitem{YF04} H. Yu and L.H. Ford, Phys. Rev. D {\bf 70}
 065009 (2004), quant-ph/0406122.



\bibitem{FR05} L.H. Ford and T.A. Roman, Phys. Rev. D {\bf 72}, 105010
(2005), gr-qc/0506026. 

\bibitem{F06}  L.H. Ford, Int. J. Theor. Phys., in press, quant-ph/0601112.

\bibitem{FW03}  L.H. Ford and C.-H. Wu, Int. J. Theor. Phys. {\bf 42},
  15 (2003).

\bibitem{MV99} R. Martin and E. Verdaguer, Phys. Rev. D {\bf 60},
  084008 (1999).

\bibitem{BF04a} J. Borgman and L.H. Ford, Phys. Rev. D {\bf 70}
 064032 (2004), gr-qc/0307043.


\bibitem{Moffat} J.W. Moffat, Phys. Rev. D {\bf 56}, 6264 (1997), 
gr-qc/9610067.

\bibitem{RF06} R.T. Thompson and L.H. Ford,  Phys. Rev. D {\bf 74},
024012 (2006), gr-qc/0601137. 

\bibitem{WF99} C.-H. Wu and L.H. Ford, Phys. Rev. D {\bf 60}, 
104013 (1999), gr-qc/9905012.

\bibitem{HR07}  B.L. Hu and A. Roura, arXiv:0708.3046. 

\bibitem{FS97}  L.H. Ford and N.F. Svaiter,  Phys. Rev. D {\bf 56},
2226 (1997), quant-ph/9704050.


\bibitem{BFP00}  C. Barrabes, V. Frolov, and R. Parentani,
  Phys. Rev. D {\bf 62}, 044020 (2000).  gr-qc/0001102.
 
\bibitem{WNF07} C.-H. Wu, K.-W. Ng, and L.H. Ford, Phys. Rev. D {\bf
  75}, 103502 (2007), gr-qc/0608002. 


\bibitem{Hawking66} S.W. Hawking, Astrophys. J. {\bf 145}, 544 (1966).

\bibitem{Olson} D.W. Olson,  Phys. Rev. D {\bf 14}, 327 (1976).





\end{thebibliography}
\end{document}